\documentclass[preprint,showpacs,preprintnumbers,amsmath,amssymb]{revtex4}

\usepackage{graphicx}% Include figure files
\usepackage{dcolumn}% Align table columns on decimal point
\usepackage{bm}% bold math

\begin{document}

\preprint{}

\title{Analytic Fits to Separable Volumes and Probabilities for Qubit-Qubit
and Qubit-Qutrit Systems}

\author{Paul B. Slater}% 
\email{slater@kitp.ucsb.edu}
\affiliation{%
ISBER, University of California, Santa Barbara, CA 93106\\
}%
\date{\today}% It is always \today, today,
             %  but any date may be explicitly specified

\begin{abstract}
We investigate the possibility of deriving analytical formulas for
the $15$-dimensional {\it separable} volumes, in terms of any of a 
number of metrics of interest (Hilbert-Schmidt [HS], Bures,...), 
of the two-qubit (four-level) systems. This would appear to 
require $15$-fold symbolic integrations over a complicated
convex body (defined by {\it both} separability and feasibility 
constraints). The associated 15-dimensional 
integrands --- in terms of the Tilma-Byrd-Sudarshan Euler-angle-based 
parameterization of the $4 \times 4$ density matrices $\rho$ ({\it J. Phys. A} 
35 [2002], 10445) --- would be 
the {\it products} of 12-dimensional Haar measure $\mu_{Haar}$
({\it common} to each metric) and 
3-dimensional measures $\mu_{metric}$ ({\it specific} 
to each metric) over the $3d$-simplex formed by the four eigenvalues 
of $\rho$.
We attempt here to estimate/determine 
the 3-dimensional integrands (the products of the various [known]
$\mu_{metric}$'s and an 
{\it unknown} symmetric
weighting function $W$) 
remaining after the (putative) 12-fold integration of $\mu_{Haar}$ over 
the twelve Euler angles. We do this by {\it first} fitting $W$, so 
that the conjectured 
HS
separable volumes and hyperareas 
({\it Phys. Rev. A} 71 [2005], 052319; cf. quant-ph/0609006) are reproduced. 
We further evaluate a number of possible such
choices of $W$ by seeing how well they {\it also} yield
the conjectured separable volumes
for the Bures, Kubo-Mori, Wigner-Yanase and (arithmetic) 
average monotone metrics and the conjectured 
separable Bures {\it hyperarea} ({\it J. Geom. Phys.} 53 [2005], 74, Table VI).
We, in fact, find two such exact (rather similar) 
choices for $W$ that give these 
five conjectured ({\it non}-HS) 
values all within 5\%. In addition to the above-mentioned Euler angle
parameterization of $\rho$, 
we make extensive use of the Bloore parameterization
({\it J. Phys. A} 9 [1976], 2059) in a companion set of two-qubit
separability analyses.

\end{abstract}

% PACS, the Physics and Astronomy
                             % Classification Scheme.
%\keywords{Suggested keywords}%Use showkeys class option if keyword
                              %display desired
\maketitle
\section{Introduction}
In a pair of major, skillful papers, making use of the theory of
 random matrices \cite{random}, Sommers and \.Zyczkowski were able to
derive explicit formulas for the volumes occupied by the 
$d= (n^2-1)$-dimensional convex set of $n \times n$ (complex) 
density matrices (as well as the $d=\frac{(n-1)(n+2)}{2}$-dimensional 
convex set of real $n \times n$ density matrices),
{\it both} in terms of the Hilbert-Schmidt (HS) metric \cite{szHS} --- inducing the flat, Euclidean geometry --- and 
the Bures metric \cite{szBures} (cf. \cite{szMore}).
(These results are also more lately discussed in the highly comprehensive
new text of Bengtsson and \.Zyczkowski \cite[chap. 14]{ingemar}.)
Of course, it would be of obvious 
considerable quantum-information-theoretic 
interest in the cases that $n$ is a {\it composite}
number, to also obtain HS and Bures volume 
formulas {\it restricted} to those states that
are {\it separable} --- the sum of {\it 
product} states --- in terms of some factorization of $n$ \cite{ZHSL}. 
Then, by taking ratios --- employing these Sommers-\.Zyczkowski 
results --- one would obtain corresponding 
separability {\it probabilities}.

In particular, again for the 15-dimensional complex 
case, $n=4 = 2 \times 2$, {\it numerical}
evidence has been adduced 
that the Bures volume of separable states is
(quite elegantly) $2^{-15} (\frac{\sqrt{2}-1}{3}) \approx 
4.2136  \cdot 10^{-6}$ \cite[Table VI]{slaterJGP} 
and the HS volume
$(5 \sqrt{3})^{-7} 
\approx 2.73707 \cdot 10^{-7}$ \cite[eq. (41)]{slaterPRA}. 
Then, taking ratios (using the Sommers-\.Zyczkowski results 
\cite{szHS,szBures}), 
we have the {\it derived} conjectures that the Bures {\it separability}
 probability
is $\frac{1680 (\sqrt{2}-1)}{\pi^8} \approx 0.0733389$ 
and the HS one, 
considerably larger, $\frac{2^2 \cdot 3 \cdot 7^2 \cdot 11 \cdot 13 \sqrt{3}}{5^4 \pi^{6}} \approx 0.242379$ \cite[eq. (43), but misprinted as $5^3$ not 
$5^4$ there]{slaterPRA}.
(Szarek, Bengtsson and \.Zyczkowski --- motivated by the numerical 
findings of \cite{slaterPRA,slaterChinese} --- have recently 
formally demonstrated 
 ``that the probability to find a random state to be separable equals 2 times the probability to find a random boundary state to be separable, provided the random states are generated uniformly with respect to the Hilbert-Schmidt (Euclidean) distance. An analogous property holds for the set of positive-partial-transpose states for an arbitrary bipartite system'' \cite{sbz} 
(cf. \cite{innami}). 
(``Since our reasoning hinges directly on the 
Euclidean geometry, 
it does not allow to predict any values of analogous ratios 
computed with respect to Bures measure, nor other measures'' 
\cite[p. L125]{sbz}.) 
These three authors 
also noted \cite[p. L125]{sbz} that ``one could try to obtain similar
results for a general class of multipartite systems''. In this 
latter vein,
numerical analyses of ours give some [but certainly not yet conclusive] 
indication that for the {\it three}-qubit
{\it tri}separable states, there is an analogous probability ratio of
6 --- rather than 2.)

However, the analytical derivation of (conjecturally) {\it exact}
formulas for these HS and Bures (as well as other, such as the Kubo-Mori
\cite{petz1994} and Wigner-Yanase 
\cite{wigneryanase,slaterPRA}) {\it separable} volumes still appears to be quite remote (cf. \cite{mathscinet}) --- the only such 
progress to report so far being 
certain exact formulas
when the number of dimensions of the 15-dimensional space of $4 \times 4$ 
density matrices has been severely 
curtailed (nullifying or holding 
constant {\it most} of the 15 parameters) to $d \leq 3$ 
\cite{pbsJak,pbsCanosa} (cf. \cite{slaterC}).
Most notably, in this research direction,
in \cite[Fig. 11]{pbsCanosa}, we were able to find 
a highly interesting/intricate (one-dimensional) continuum 
($-\infty < \beta <\infty$) of two-dimensional 
(the associated 
parameters being $b_{1}$, the {\it mean}, and $\sigma_{q}^2$,   
the {\it variance} of the Bell-CHSH observable) 
HS separability
probabilities, in which the {\it golden ratio} \cite{livio} was 
featured, among other items. (The associated 
HS volume element --- $\frac{1}{32 \beta (1+\beta)} 
d \beta d b_{q} d \sigma^2_{q}$ --- is
{\it independent} of $b_{1}$ and $\sigma_{q}^2$ in this 
three-dimensional scenario. Extensions to higher-dimensional scenarios 
$d>3$ appear problematical, though.)
Further, in \cite{pbsJak}, building upon work of 
Jak\'obczyk and Siennicki \cite{jak}, we obtained a 
remarkably wide-ranging variety of exact HS 
separability ($n=4, 6$) and PPT (positive partial transpose) 
($n=8, 9, 10$) probabilities based on 
{\it two}-dimensional {\it sections} of sets of 
(generalized) Bloch vectors corresponding to $n \times n$ 
density matrices.

In this paper we are able to report some 
additional progress in these directions.
We obtain exact formulas for certain $d=4, n =4$ scenarios and 
{\it upper} bounds for $d=7$ and $d=9$ instances.
(Nevertheless, 
the full $d=9$ and/or $d =15$, $n=4$ real and complex scenarios
still appear quite daunting --- due to the numerous
separability constraints at work, some being active [binding] 
in certain regions and 
in complementary regions, inactive [nonbinding]. 
``The geometry of the $15$-dimensional set of separable states of two
qubits is not easy to describe'' \cite[p. L125]{sbz}.)

To proceed initially 
(secs.~\ref{sc1} \ref{sc2} \ref{sc3}), 
we employ the (quite simple) form of parameterization of the density matrices
put forth by Bloore \cite{bloore} some thirty years ago. 
(Of course, there are several 
other possible parametrizations 
\cite{kk,byrd,sudarshan,vanik,fano,scutaru,stan}, a number of 
which we have also utilized in various studies \cite{slaterA,slaterqip} 
to estimate volumes of 
separable states. Our greatest progress at this stage, 
in terms of increasing dimensionality,  has been achieved
with the Bloore parameterization --- due to a certain 
computationally attractive feature of it, allowing us to 
{\it decouple} diagonal and non-diagonal parameters --- as detailed 
shortly below.)

In our final 
(quite differently structured) series of analyses (sec.~\ref{weighting functionsec}), though, we employ not the Bloore
parameterization, but the Euler-angle-based one 
of Tilma, Byrd and Sudarshan \cite{sudarshan}.
Our motivation here is to bypass/circumvent the necessity of 
the putatively achieveable, but computationally daunting 
first twelve steps (over the twelve Euler angles) of a 15-fold
integration. 
(``we would like to derive a subset of the ranges of the Euler angle 
parameters...dividing the 15-parameter space into entangled and 
separable subsets. Unfortunately, due to the complicated nature of the parameterization, 
both numerical and symbolic calculations of the eigenvalues of the 
partial transpose...have become computationally intractable using
standard mathematical software'' \cite[p. 10453]{sudarshan}.)
We seek to find the {\it three}-dimensional (weighting) function 
$W(\lambda_{1},\lambda_{2},\lambda_{3},\lambda_{4})$
of the {\it four} 
eigenvalues ($\lambda_{4}=1-\lambda_{1}-\lambda_{2}-\lambda_{3}$) 
which would yield our various {\it conjectured} 15-dimensional
separable volumes and 14-dimensional separable hyperareas 
for the two-qubit systems.
\cite{slaterJGP,slaterPRA}.
In sec.~\ref{n6}, we also apply this approach 
exploratorily to the $n=6, d=35$ case
of qubit-{\it qutrit} pairs, seeking to find a suitable 
5-dimensional weighting function 
$W_{n=6}(\lambda_{1},\lambda_{2},\lambda_{3},\lambda_{4},\lambda_{5},\lambda_{6})$.

Let us also point out to the reader, our even more recent 
companion-type study
\cite{slaterHSPRA}, in which solely for the 
(relatively simple) Hilbert-Schmidt (non-monotone)
metric, are we able (making use of the Bloore parameterization) 
to reduce the two-qubit (both real and complex)  
separable volume determination
problems to those of finding  {\it one}-dimensional weighting functions
(which turned out to be well-approximated by 
certain incomplete beta functions --- functions of the ratio
$\frac{\rho_{11} \rho_{44}}{\rho_{22} \rho_{33}}$ of diagonal entries 
of the $4 \times 4$ density matrices $\rho$).
\section{Bloore parameterization of the density matrices} \label{sc1}
The main presentation of Bloore \cite{bloore} 
was made in terms of the $3 \times 3$ ($n=3$)
density matrices. It is clearly easily extendible to cases
$n >  3$.
The fundamental idea is to {\it scale} the {\it 
off}-diagonal elements $(\rho_{ij}, 
i \neq j)$ 
of the density matrix in terms of the {\it square roots} of the diagonal 
entries ($\rho_{ii}$). That is, we set (introducing the new [Bloore] 
variables $z_{ij}$),
\begin{equation}
\rho_{ij} = \sqrt{\rho_{ii} 
\rho_{jj}} z_{ij}.
\end{equation}
 This allows the {\it determinant} of $\rho$ (and analogously 
all its
{\it principal minors}) to be expressible as the {\it product} 
($|\rho| = A B$) of
two factors, one ($A$)  of which is itself simply the product of 
(positive) 
diagonal entries ($\rho_{ii}$) 
and the other --- in the $n=4$ case under investigation \newline 
here (easily extendible from the case of real density matrices 
to complex ones) --- 
\begin{equation} \label{B}
B= \left(z_{34}^2-1\right) z_{12}^2+2 \left(z_{14}
   \left(z_{24}-z_{23} z_{34}\right)+z_{13}
   \left(z_{23}-z_{24} z_{34}\right)\right)
   z_{12}-z_{23}^2-z_{24}^2-z_{34}^2+
\end{equation}
\begin{displaymath}
z_{14}^2
   \left(z_{23}^2-1\right)+ z_{13}^2
   \left(z_{24}^2-1\right)+2 z_{23} z_{24} z_{34}+2 z_{13}
   z_{14} \left(z_{34}-z_{23} z_{24}\right)+1,
\end{displaymath}
involving ({\it only}) the $z_{ij}$'s ($i \neq j$) 
\cite[eqs. (15), (17)]{bloore}.
Since, clearly, the factor $A$ is positive in all nondegenerate cases 
($\rho_{ii} \geq 0$),
one can --- by {\it only} analyzing $B$ --- essentially 
ignore the diagonal entries (and thus {\it reduce} by ($n-1$)) the
dimensionality of the problem of finding nonnegativity 
conditions to impose on $\rho$.
This is the feature we will seek to maximally 
exploit here.

It is, of course, necessary and sufficient for $\rho$ to serve
as a density matrix (that is, an Hermitian, nonnegative definite, trace
one matrix) that {\it all} its {\it principal} minors be nonnegative 
\cite{horn}.
The condition --- quite natural in the Bloore 
parameterization --- that all the principal $2 \times 2$ minors be
nonnegative requires simply that $-1 \leq z_{ij} \leq 1, i \neq j$. The 
{\it joint} conditions that
{\it all} the principal minors be nonnegative are not as
readily apparent. But for the 9-dimensional {\it real} 
case $n=4$ --- that is, $\Im(\rho_{ij})=0$ --- we have been able to obtain 
one such set,
using the Mathematica implementation of the {\it cylindrical
algorithm decomposition} \cite{cylindrical}.
(The set of solutions of any system of real algebraic equations
and inequalities can be decomposed into a finite number of
``cylindrical'' parts \cite{strzebonski}.)
Applying it, we were able to express the 
conditions that 
an arbitrary  9-dimensional $4 \times 4$ real density matrix 
$\rho$ must fulfill.
These took the form, $z_{12}, z_{13}, z_{14} \in [-1,1]$ and 
\begin{equation} \label{limits}
 z_{23} \in [Z^-_{23},Z^+_{23}], \hspace{.1in}
 z_{24} \in [Z^-_{24},Z^+_{24}], \hspace{.1in}
 z_{34} \in [Z^-_{34},Z^+_{34}],
\end{equation}
where
\begin{equation}
Z^{\pm}_{23} =z_{12} z_{13} \pm \sqrt{1-z_{12}^2} \sqrt{1-z_{13}^2} , 
\hspace{.1in}
Z^{\pm}_{24} =z_{12} z_{14} \pm \sqrt{1-z_{12}^2} \sqrt{1-z_{14}^2} ,
\end{equation}
\begin{displaymath}
Z^{\pm}_{34} = \frac{z_{13} z_{14} -z_{12} z_{14} z_{23} -z_{12} z_{13} z_{24} +z_{23}
z_{24} \pm s}{1-z_{12}^2},
\end{displaymath}
and
\begin{equation}
s = \sqrt{-1 +z_{12}^2 +z_{13}^2 -2 z_{12} z_{13} z_{23} +z_{23}^2}
\sqrt{-1 +z_{12}^2 +z_{14}^2 -2 z_{12} z_{14} z_{24} +z_{24}^2}.
\end{equation}
Making use of these results, we were able to confirm {\it via} exact 
{\it symbolic} integrations, 
the (formally demonstrated) 
result of \.Zyczkowski and Sommers
\cite{szHS} that the HS volume of the {\it real} 
two-qubit ($n=4$) states is
$\frac{\pi^4}{60480} \approx 0.0016106$.
(This result was also achievable through a somewhat different
Mathematica computation, using the {\it implicit} integration feature
first 
introduced in version 5.1. That is, the only integration limits employed were
that $z_{ij} \in [-1,1], i \neq j$ --- {\it broader} than those in 
(\ref{limits}) --- while the Boolean constraints were imposed that 
the determinant of $\rho$ and {\it one} [all that is needed to ensure
nonnegativity] of its principal $3 \times 3$ minors be nonnegative.)

However, when we tried to combine these integration limits (\ref{limits}) 
with
the (Peres-Horodecki \cite{asher,michal,bruss} $n=4$) 
{\it separability} constraint that the determinant ($C =|\rho_{PT}|$) 
of the partial
transpose of $\rho$ be nonnegative \cite[Thm. 5]{ver}, 
we {\it exceeded} the memory availabilities of our workstations.
In general, the term $C$ --- unlike the earlier term $B$ --- unavoidably 
involves 
the diagonal entries ($\rho_{ii}$), so the
dimension of the accompanying integration problems must {\it increase} --- 
in the $9$-dimensional real  $n=4$ case from 6 to 9.
\subsection{Restricting diagonal entries}
Nevertheless, we found that by imposing the condition that the four diagonal
entries ($\rho_{ii}, i =1,\ldots,4$) fall into two equal pairs
(say, $\rho_{11} =\rho_{22}$ and $\rho_{33}=\rho_{44}$, so that 
$\rho_{33} =\frac{(1- 2 \rho_{11})}{2}$), the determinant ($C$) of 
the partial transpose could now be expressed as the {\it product}
\begin{equation}
C = |\rho_{PT}| = \frac{1}{4} (1-2 \rho_{11})^2 \rho_{11}^2 D, 
\hspace{.5in} 0\leq \rho_{11} \leq  \frac{1}{2},
\end{equation}
where
\begin{equation} \label{D}
D=\left(z_{24}^2-1\right) z_{13}^2+2 z_{23}
   \left(z_{34}-z_{14} z_{24}\right)
   z_{13}-z_{24}^2-z_{34}^2+
\end{equation}
\begin{displaymath}
\left(z_{14}-1\right)
   \left(z_{14}+1\right) \left(z_{23}-1\right)
   \left(z_{23}+1\right)+2 z_{14} z_{24} z_{34}+
\end{displaymath}
\begin{displaymath}
\left(z_{34}^2-1\right) z_{12}^2+2 \left(z_{13}
   z_{14}+z_{23} z_{24}-\left(z_{14} z_{23}+z_{13}
   z_{24}\right) z_{34}\right) z_{12}.
\end{displaymath}
Thus, the term $D$ (like $B$ in (\ref{B})) 
is itself {\it independent} of the diagonal entries
of $\rho$ (in particular, the specific value of $\rho_{11}$) --- allowing us to proceed, as indicated, 
with integrations in a {\it lower}-dimensional ($d \leq 7$) setting. 
(This same form of factorizability takes place, as well, for all the [four]
$3 \times 3$ and [six] $2 \times 2$ minors of $\rho_{PT}$.)
If we can further guarantee the nonnegativity of $D$ --- in addition 
to that of $B$  --- we can ensure
separability of $\rho$. Let us also note that
\begin{equation} \label{factors}
B-D= 2 (z_{14} -z_{23}) (z_{13} -z_{24}) (z_{12} -z_{34}).
\end{equation}
(So, if any of the three factors in (\ref{factors}) 
are zero, the associated
state must be separable.)

We will now proceed
to some specific analyses within this more restrictive framework 
(7-dimensional in nature, since we started with the 9-dimensional real
setting and have essentially only one free diagonal parameter [$\rho_{11}$] 
left).
\section{7-Dimensional Real Setting
($\rho_{11}=\rho_{22},\rho_{33}=\rho_{44}$)} \label{sc2}
\subsection{7-dimensional analysis}
The associated $7$-dimensional HS volume of all these states
(separable and nonseparable) is $\frac{\pi^2}{15120} \approx
0.000652752$. The six principal $2 \times 2$ minors of the partial transpose
simply yield the (Bloore) 
conditions that $z_{ij} \in [-1,1]$, so nothing can be
gained --- in terms of obtaining upper bounds on the separable volume --- by 
using the nonnegativity of these six minors as further constraints in 
our integrations. However,
if we require that {\it one} of the $3 \times 3$ principal minors of the 
partial transpose be nonnegative, 
we do succeed in obtaining a nontrivial upper bound of
$\frac{\pi^4}{172032} \approx 0.000566227$ on the 
$7$-dimensional volume of 
separable states. So, we have a derived {\it upper bound} 
(probably rather weak, we surmise) 
on the HS separability
probability for our $7$-dimensional real set of $4 \times 4$ 
density matrices of $\frac{45 \pi^2}{512} \approx 0.867446$.
\subsection{4-dimensional analyses}
\subsubsection{$z_{12}=z_{23}=z_{24}=0$}
Here, we set the three indicated Bloore  parameters to
zero. (So, the {\it four} free parameters of the initial {\it seven}
are $\rho_{11}, z_{13}, z_{14}$ 
and $z_{34}$.)
Then, using the implicit integration feature of Mathematica
(rather than the limits (\ref{limits})),  we 
were able to obtain for the total HS volume
$\frac{\pi^2}{384} \approx 0.0257021$ and (further 
{\it adding} the separability constraint that
$D$, given by (\ref{D}),  be greater than 0) the separable volume of 
$\frac{4 + \pi^2}{1536} \approx 0.00902969$.
Taking the appropriate quotient, we find for the HS separability
probability for this scenario (quite elegantly),
$\frac{4 + \pi^2}{4 \pi^2} \approx 0.351321$.
\subsubsection{$z_{23}=0,z_{24}=0,z_{34}=0$}
Now, $HS_{vol}^{tot} = \frac{\pi}{144} \approx 0.0218166$ and
$HS_{vol}^{sep}= \frac{4 + \pi^2}{1536} \approx 
0.00902969$, 
so \newline $HS_{sep prob} \equiv \frac{HS_{vol}^{sep}}{HS_{vol}^{tot}} = 
\frac{3 (4 + \pi^2)}{32 \pi} \approx 0.413891$.

There are twenty possible $4$-dimensional scenarios. Of these four each
have one of these two nontrivial HS separability probabilities 
($\frac{4 + \pi^2}{4 \pi^2}$ or 
$\frac{3 (4 + \pi^2)}{32 \pi}$). For
the remaining twelve cases, 
the HS separability probabilities are simply 1.
\subsection{3-dimensional analyses}
We also observe that for the fifteen possible $3$-dimensional
scenarios, all the HS separability probabilities are 
(trivially) 1. (Eight of these
scenarios have HS total volume equal to $\frac{\pi^2}{128}$, four, 
$\frac{\pi}{48}$ and three $\frac{1}{12}$.)
\subsection{$5$-dimensional analyses}
For all twelve possible scenarios, setting two $z_{ij}$'s to zero, 
the total HS volume of states is
$\frac{\pi^2}{1440} \approx 0.00685389$.
\subsubsection{$z_{23}=z_{24}=0$}
To compute the separable HS volume we had to resort to
numerical means, obtaining 0.00532303, for a separability probability
of 0.776643. (This is also the probability for 
the 5-dimensional scenario $z_{12}=z_{13}=0$ --- and [at least] five others.)
\subsubsection{$z_{14}=0=z_{23}=0$}
Here, as indicated, 
the total HS volume is $\frac{\pi^2}{1440} \approx 0.00685389$.
This is also the separable volume  --- since $B=D$ in this case. 
(This is also the situtation with the scenarios $z_{13}=z_{24}=0$ and
$z_{12}=z_{34}=0$.)
\section{$9$-dimensional real case} \label{sc3}
As previously mentioned, we know from the Sommers-\.Zyczkowski 
analyses \cite{szHS} that the
HS volume of the $9$-dimensional convex set of {\it real}
$4 \times 4$ density matrices is $\frac{\pi^4}{60480} \approx 
0.0016106$. (In this section --- to fully accord with 
their results \cite{szHS} --- we 
have to adjust by an overall scaling factor of $2^4 =16$ the results of 
our usual integration procedure employed previously 
in this study, in which we simply employed 1 as our integrand, rather
than some other constant. This scaling, 
of course, does not have any impact on
separability {\it probabilities}.)
We can also computationally verify the Sommers-\.Zyczkowski HS
volume formula in either of two manners:
(a) employing the integration limits (\ref{limits}) 
obtained by application of
the cylindrical decomposition algorithm or (b) implementing the
implicit integration feature (requiring here that the determinant of
$\rho$ and one of its $3 \times 3$ principal minors be nonnegative) 
of Mathematica, using for the integration limits
simply that $-1 \leq z_{ij} \leq 1$ for all $i \neq j$.
We investigated 
how far we could  proceed in the full $9$-dimensional $n=4$ real case, 
by imposing increasingly greater requirements (corresponding to the 
Peres-Horodecki criterion \cite{asher,michal}) that would need to 
be fulfilled for $\rho_{PT}$ to be nonnegative definite.
Now, of the six $2 \times 2$ principal minors of $\rho_{PT}$, only two are
distinct from the six such minors of $\rho$ itself.
If we demand that, in addition, to the 
feasibility constraints on $\rho$ that {\it 
one} of
these two minors be nonnegative we
obtain 
(to high precision) 
0.0014242052589  yielding an {\it upper bound} on the HS separability
probability of the $9$-dimensional real two-qubit states of
0.88426997055. Further (tighter) numerical results --- using 
additional principal minors of $\rho_{PT}$ (even the 
remaining $2 \times 2$ nonredundant one)--- proved difficult to
achieve, however.
\section{$3$-Dimensional weight for the 
{\it Full} $15$-Dimensional Problem} \label{weighting functionsec}
The direct straightforward (``brute force'') computations 
by {\it symbolic} means of the $15$-dimensional 
{\it separable} 
volumes (for any of a wide variety 
of metrics) of the (complex) two-qubit states 
seem to far exceed present workstation 
capabilities. Nevertheless, we will here try to gain
some {\it analytical} insight into these formidable problems.

Let us first note that the computation of the separable volumes 
could be seen to require the
evaluation (in which the Peres-Horodecki separability 
and feasibility criteria are {\it both}
 enforced) 
of a $15$-fold integral. Following the innovative Euler-angle-based 
parameterization of $4 \times 4$ density matrices of 
Tilma, Byrd and Sudarshan \cite{sudarshan}, 
the first twelve variables over which to integrate can be taken 
to be the twelve Euler angles ($\alpha_{i}, i = 1,...,12$) and the last
three to be the eigenvalues $\lambda_{1}, \lambda_{2}, \lambda_{3}$ of
the density matrix $\rho$. 
(Of course, $\Sigma_{i=1}^{4} \lambda_{i} =1$.) 
Now for any of the metrics of interest, the 
associated $15$-dimensional integrand
can be represented as the {\it product} of 
a 12-dimensional Haar measure ($\mu_{Haar}$) \cite[eq. (34)]{sudarshan}
 ({\it common} to {\it all} the metrics 
of interest) over the twelve Euler angles 
\begin{equation}
\mu_{Haar} =\cos(\alpha_{4})^3 \cos (\alpha_{6}) 
\cos(\alpha_{10}) 
\sin(2 \alpha_{2}) \sin(\alpha_{4}) \sin(\alpha_{6})^5 \sin(2 \alpha_{8}) 
\sin(\alpha_{10})^3 \sin(2 \alpha_{12}) 
d \alpha_{12}...d \alpha_{1}
\end{equation}
($0 \leq \alpha_{even} \leq \frac{\pi}{2}, 0 \leq \alpha_{odd} \leq \pi$ 
\cite[eq. (47)]{sudarshan}) 
and a 3-dimensional {\it metric-specific} measure ($\mu_{metric}$) 
over the eigenvalues (cf. \cite{hall}). (So, if the Peres-Horodecki criterion
is {\it not} enforced, we simply obtain the [known in the Bures and 
HS cases] volumes of the separable
{\it and} nonseparable states, the volumes
decomposing into the {\it products} of the 
results of 3-fold and 12-fold integrations 
\cite[eq. (3.7)]{szHS},\cite[eq. (3.17)]{szBures}.)

Consequently, after the first twelve steps of the (presumptively 
theoretically achieveable, but apparently 
totally impractical) integrations for the separable 
volumes, we can imagine obtaining  {\it three}-dimensional
integration problems (over the three-dimensional simplex of 
eigenvalues). The integrands of the problem would now be the {\it products}
of $\mu_{metric}$ and a {\it common} 
three-dimensional {\it weighting function} $W$, 
acquired during the course of the 12-fold integration. 
Certainly, $W$ should 
be a {\it symmetric} function \cite{ig} of the four eigenvalues.
\subsection{First analysis} \label{firstanalysis}
We will try to fit $W$ to our various conjectures  
for the separable volumes \cite{slaterJGP,slaterPRA}, 
previously obtained by {\it numerical} 
methods. 
In particular, we have found (after some limited trial and error) that
the (symmetric) choice (being neither a {\it convex} nor a {\it concave}
 function,  we observed --- nor even approximately so),
\begin{equation} \label{weighting function}
W(\lambda_{1},\lambda_{2},\lambda_{3},\lambda_{4}) 
= 6086.  (\lambda_{1} \lambda_{2} \lambda_{3} + 
\lambda_{1} \lambda_{2} \lambda_{4} +\lambda_{1} \lambda_{3} \lambda_{4} +\lambda_{2} \lambda_{3} \lambda_{4})^{\frac{53}{20}},
\end{equation}
reproduces the conjectures for {\it both} 
the {\it Hilbert-Schmidt} volumes and hyperareas
of the separable two-qubit states to good accuracy (0.01643\%). 
(These conjectured volume 
and hyperarea are $(5 \sqrt{3})^{-7} \approx 
2.73707 \cdot 10^{-7}$ and 
$(3^2 5^6)^{-1} 
\approx 7.11111 \cdot 10^{-6}$, respectively
 \cite[eqs. (41), (42)]{slaterPRA}.)
For the computation of the 14-dimensional separable hyperarea, the weighting function 
(\ref{weighting function}) reduces 
(since we can take $\lambda_{4} = 0$) to
\begin{equation} \label{weighting function2}
W(\lambda_{1},\lambda_{2},\lambda_{3})
= 6086 (\lambda_{1} \lambda_{2} \lambda_{3} )^{\frac{53}{20}}.
\end{equation}
Now, the {\it acid} test of the legitimacy/validity of our choice of 
weighting function --- and the {\it raison d'\^etre} of our exercise --- is 
to see how well (in {\it addition} to the separable 
Hilbert-Schmidt volumes and hyperareas (which we constructed to
satisfy the Szarek-Bengtsson-\.Zyczkowski {\it two}-fold 
ratio \cite{sbz}) it reproduces (our presumptively correct) 
conjectures for metrics {\it other} than the Hilbert-Schmidt one.

For the Bures (minimal monotone \cite{hansen}) 
metric, we found that the use of the weighting function (\ref{weighting function}) --- coupled with
$\mu_{Bures}$ --- predicted  a 
separable volume 0.938275 times the magnitude of the 
conjectured value of $2^{-15} (\frac{\sqrt{2}-1}{3}) \approx 4.2136  
\cdot 10^{-6}$. 
For the Kubo-Mori monotone metric, we obtained an estimate that is 0.910768 
times the magnitude of the conjectured value of $2^{-15}(10 (\sqrt{2}-1)) 
\approx 0.000126408$, for the (arithmetic) {\it average} monotone 
metric, 0.903281 times the magnitude of the conjectured value of
$2^{-15} (\frac{29}{9} (\sqrt{2}-1)) \approx 0.0000407314$, and for 
the Wigner-Yanase monotone metric, 0.919585 times the conjectured value
of $2^{-15} (\frac{7}{4} (\sqrt{2}-1)) \approx 0.000221214$.
So, our choice of $W$ works rather well, at least for a first 
simply heuristic effort. (In \cite{slaterJGP}, we also had additional 
volume 
conjectures for the GKS (Grosse-Krattenthaler-Slater) (``quasi-Bures'') 
monotone metric (cf. \cite{chernoff}). However, we encountered numerical difficulties in trying
to analyze it here, in the fashion of the other metrics.)

If we try, as well, 
to predict the 14-dimensional separable {\it hyperarea} for the
Bures metric using (\ref{weighting function2}), we obtain an estimate 
of 0.0000262122, which is 0.940364 times as large as the conjectured value
of $2^{-14} (\frac{43 (\sqrt{2}-1)}{39}) \approx 0.0000278746$ 
\cite[Table VI]{slaterJGP}. 
(For the Kubo-Mori metric, we also obtain an estimate of the separable hyperarea 
of 0.0000399861 and of the separable probability of a state on the 
14-dimensional boundary 
of 0.0214689 --- but there were no prior conjectures for these
quantities in \cite[Table VI]{slaterJGP}. For the arithmetic average metric,
our estimate of the separable hyperarea is 0.738784, while our 
conjecture in \cite[Table VI]{slaterJGP} [to be corrected by a factor of 
8 as noted in \cite[p. 1-11]{slaterPRA}] amounts to 
$2^{-14} (\frac{255 (\sqrt{2}-1)}{128}) \approx 0.825191$.)

\subsection{Further analyses --- exact weighting functions}
We, then, altered our analytical strategy somewhat and succeeded (twice)
in reproducing our five indicators ---- the Bures, Kubo-Mori, Wigner-Yanase 
and (arithmetic) average separable volumes and the Bures hyperarea --- 
all now within 5\% of their conjectured values. We accomplished this by 
{\it exactly}
fitting (by finding the values for $a$ and $b$) 
the conjectured HS separable volume and separable hyperarea to
weighting functions of the form (and their $\lambda_{4}=0$ reductions),
\begin{equation} \label{exactexact}
W(\lambda_{1},\lambda_{2},\lambda_{3},\lambda_{4}) = a (\Sigma^{4}_{i<j} \lambda_{i} \lambda_{j})^{m_{1}} 
+ b (\Sigma^{4}_{i<j<k} \lambda_{i} \lambda_{j} \lambda_{k})^{m_{2}}.
\end{equation}
We conducted separate analyses for pairs of 
low integral values of 
the exponents ($1 \leq m_{1}, m_{2} \leq 4$).
For $m_{1} =3, m_{2}=3$ we had the results
\begin{equation} \label{EXACT1}
a= \frac{325909584 \sqrt{3}}{464375 \pi ^6} \approx 1.26422, 
b= \frac{5070990172248 \sqrt{3}}{464375 \pi ^6} \approx 19673.7
\end{equation}
$(\frac{b}{a} =\frac{31119}{2})$,
and for $m_{1} =4, m_{2}=3$,
\begin{equation} \label{EXACT2}
a= \frac{8834477652 \sqrt{3}}{3109375 \pi ^6} \approx 5.11881, 
b = \frac{33503284082268 \sqrt{3}}{3109375 \pi ^6} \approx 19412.2
\end{equation}
$(\frac{b}{a}=\frac{11377}{3})$.
(Numerical tests showed that neither of these two functions was
Schur-convex nor Schur-concave \cite{nielsen}, nor even approximately so.)
As indicated, for both these settings of (\ref{exactexact}), our five 
({\it non}-HS)
indicators all lay within {\it at most} 5\% of their conjectured values 
\cite{slaterJGP}.
(Additionally, the estimated arithmetic average separable hyperarea 
lay within 5\% of the conjectured value for the $m_{1}=4,m_{2}=3$ case, 
and 10\% for the other. Since the two functions {\it both} fit the conjectured  separable HS volume and hyperarea, any linear combination of them
will also. We have found that by weighting the function associated 
with (\ref{EXACT1}) by 0.570347, and the other function by 
0.429653, there is no deviation in the 
associated non-HS indicators by more than 
3.61\%.) The closeness of our estimates to their conjectured
values, certainly it would seem, should lend some further support 
(beyond the original numerical evidence \cite{slaterJGP,slaterPRA}) for the 
reasonableness of the associated conjectures.

Of course, it behooves us and possibly other interested 
researchers, at this stage, to explore the properties 
(and desirably derive {\it guiding} principles) of 
additional candidates
for the presumptive three-dimensional weighting function $W$.
(We are compelled to note, however, that contrary to our
construction of $W$ so far, that it is clear that $W$ has to be 
simple {\it flat}
in some finite neighborhood 
of the fully mixed state $\lambda_{1}=\lambda_{2}=\lambda_{3}
=\lambda_{4}= \frac{1}{4}$ \cite{sepsize1} \cite[sec. 15.5]{ingemar}. 
So, it seems that $W$ needs to be defined in a {\it piecemeal} manner over 
differing domains (cf. \cite{pbsCanosa}).)

\section{5-dimensional qubit-{\it qutrit} weighting function} \label{n6}
Proceeding analogously to our (first) 
analysis in sec.~\ref{firstanalysis} 
for the two-qubit
case ($n=4$), we sought to obtain a 5-dimensional weighting function $W_{n=6}$ 
for the qubit-qutrit case ($n=6$).
Using the conjectures --- based on extensive numerical results --- stated in 
\cite[sec. VI.D.2]{slaterPRA}, 
in particular, those for the {\it Hilbert-Schmidt}  separable 
volume and separable hyperarea of the
$35$-dimensional convex set of $6 \times 6$ density matrices 
(conjectured to be  $(2^{45} \cdot 3 \cdot 5^{13} \cdot 7 \sqrt{30})^{-1}
\approx 2.02423 \cdot 10^{-25}$ and  
$(2^{46} \cdot 3 \cdot 5^{12})^{-1}
\approx 1.94026 \cdot 10^{-23}$, respectively), we fitted
the weighting function (reproducing the HS separable volume to very high accuracy 
and the hyperarea to an 
accuracy of .7\%),
\begin{equation}
W_{n=6}(\lambda_{1},\lambda_{2},\lambda_{3},\lambda_{4},\lambda_{5},\lambda_{6}) =
986304. 
(\Sigma_{i<j<k<l<m}^{6} \lambda_{i} \lambda_{j} \lambda_{k} \lambda_{l} \lambda_{m})^{\frac{9}{5}}.
\end{equation}

Now, in applying this to the {\it Bures} metric, we derived an estimate 
of the 35-dimensional Bures volume that was 
1.82587 times as large as the conjectured value of $2^{-77} 
\cdot 3 \sqrt{8642986 \pi} \approx 1.03447 \cdot 10^{-19}$ 
\cite[eq. (32)]{slaterJGP}. Our estimate of the 34-dimensional Bures 
hyperarea was 1.91223 times as large as the conjectured value 
of $2^{43} \cdot 3 \cdot 5 \sqrt{8462986 \pi} \approx 
1.45449 \cdot 10^{-18}$ \cite[eq. (33)]{slaterPRA}. 
Though, somewhat disappointingly large, these two early 
results are certainly of the same order of magnitude as the conjectures
(and the underlying supporting numerical evidence), and suggest 
additional research.

\section{Remarks}
We have sought to determine a certain 3-dimensional weighting function
by fitting the conjectured values of the Hilbert-Schmidt 
separable 15-dimensional 
volume and 14-dimensional hyperarea \cite[eqs. (41), (42)]{slaterPRA}.
It would be of interest to attempt
to fit {\it additional} conjectured values as well (such as those for the
Bures [minimal monotone] metric).  (No conjecture is presently available
for the HS separable 
hyperarea of the 11-dimensional space spanned by the rank-2 density 
matrices. Otherwise it could be incorporated into our 
further analyses too --- if 
the weighting function did not degenerate with two zero eigenvalues 
present, and if the corresponding 11-dimensional 
separable hyperarea is not actually zero [cf. \cite{lockhart} \cite[sec. VI.C.4]{slaterPRA}].)

Let us --- as was done in \cite{sbz} --- bring to the reader's 
attention some other studies, such as  \cite{sz1,sz2,gb1,gb2,hildebrand} 
pertaining to volumes of sets of
separable and/or positive-partial-transpose states, as well as our
more recent analysis \cite{slaterHSPRA}, concerning the Hilbert-Schmidt
metric.
(It becomes quite clear in this last study, that the separability constraint
on two-qubit systems is, in general, {\it quartic} in nature 
(cf. \cite{wang,ulrich}), thus, 
to some extent, explaining the associated difficulties in enforcing it --- as wwell as raising certain interesting topological questions
(cf. \cite{fortuna}).)

In conclusion, 
let us also make reference to a certain capsule review \cite{mathscinet} 
in the database MathSciNet of our 
previous paper \cite{slaterJGP} in this journal. 
In particular, we add 
emphasis to the final
sentence of the review, devising a response to which 
comment has been the main
motivation of this paper, as well as that of \cite{slaterHSPRA}.

``The paper concerns properties of the convex set of 
separable two-qubit states. Although the positive partial 
transpose criterion gives in this very case a concrete answer to the 
question of whether a given mixed state is separable, 
the geometry of the 15-dimensional set ${S}$ of separable states 
is still not well understood. 

The author analyzes numerically the volume of the set ${S}$ 
with respect to measures induced by several monotone metrics. 
In particular, he studies the one-parameter family of metrics 
interpolating between the maximal and the minimal (Bures) metrics. 

Working with the Bures measure he conjectures that the relative volume of 
the set of separable states is equal to the silver mean, $\sigma=\sqrt{2}-1$. 
In a similar way the volume of the $14$-dimensional hyperarea of 
${S}$ is estimated with respect to various measures, and 
the ratios area/volume are analyzed. 

{\it The conjectures of Slater, based on numerical integration, still 
await analytical confirmation.}''

\begin{acknowledgments}
I wish to express gratitude to the Kavli Institute for Theoretical
Physics (KITP)
for computational support in this research.

\end{acknowledgments}

\bibliography{FJ5}% Produces the bibliography via BibTeX.

\begin{thebibliography}{52}
\expandafter\ifx\csname natexlab\endcsname\relax\def\natexlab#1{#1}\fi
\expandafter\ifx\csname bibnamefont\endcsname\relax
  \def\bibnamefont#1{#1}\fi
\expandafter\ifx\csname bibfnamefont\endcsname\relax
  \def\bibfnamefont#1{#1}\fi
\expandafter\ifx\csname citenamefont\endcsname\relax
  \def\citenamefont#1{#1}\fi
\expandafter\ifx\csname url\endcsname\relax
  \def\url#1{\texttt{#1}}\fi
\expandafter\ifx\csname urlprefix\endcsname\relax\def\urlprefix{URL }\fi
\providecommand{\bibinfo}[2]{#2}
\providecommand{\eprint}[2][]{\url{#2}}

\bibitem[{\citenamefont{Mehta}(2004)}]{random}
\bibinfo{author}{\bibfnamefont{M.~L.} \bibnamefont{Mehta}},
  \emph{\bibinfo{title}{Random Matrices}}
  (\bibinfo{publisher}{Elsevier/Academic}, \bibinfo{address}{Amsterdam},
  \bibinfo{year}{2004}).

\bibitem[{\citenamefont{{\.Z}yczkowski and Sommers}(2003)}]{szHS}
\bibinfo{author}{\bibfnamefont{K.}~\bibnamefont{{\.Z}yczkowski}}
  \bibnamefont{and} \bibinfo{author}{\bibfnamefont{H.-J.}
  \bibnamefont{Sommers}}, \bibinfo{journal}{J. Phys. A}
  \textbf{\bibinfo{volume}{36}}, \bibinfo{pages}{10115} (\bibinfo{year}{2003}).

\bibitem[{\citenamefont{Sommers and {\.Z}yczkowski}(2003)}]{szBures}
\bibinfo{author}{\bibfnamefont{H.-J.} \bibnamefont{Sommers}} \bibnamefont{and}
  \bibinfo{author}{\bibfnamefont{K.}~\bibnamefont{{\.Z}yczkowski}},
  \bibinfo{journal}{J. Phys. A} \textbf{\bibinfo{volume}{36}},
  \bibinfo{pages}{10083} (\bibinfo{year}{2003}).

\bibitem[{\citenamefont{Sommers and {\.Z}yczkowski}(2004)}]{szMore}
\bibinfo{author}{\bibfnamefont{H.-J.} \bibnamefont{Sommers}} \bibnamefont{and}
  \bibinfo{author}{\bibfnamefont{K.}~\bibnamefont{{\.Z}yczkowski}},
  \bibinfo{journal}{J. Phys. A} \textbf{\bibinfo{volume}{37}},
  \bibinfo{pages}{8457} (\bibinfo{year}{2004}).

\bibitem[{\citenamefont{Bengtsson and {\.Z}yczkowski}(2006)}]{ingemar}
\bibinfo{author}{\bibfnamefont{I.}~\bibnamefont{Bengtsson}} \bibnamefont{and}
  \bibinfo{author}{\bibfnamefont{K.}~\bibnamefont{{\.Z}yczkowski}},
  \emph{\bibinfo{title}{Geometry of quantum states: an introduction to quantum
  entanglement}} (\bibinfo{publisher}{Cambridge Univ. Press},
  \bibinfo{address}{Cambridge}, \bibinfo{year}{2006}).

\bibitem[{\citenamefont{{\.Z}yczkowski
  et~al.}(1998)\citenamefont{{\.Z}yczkowski, Horodecki, Sanpera, and
  Lewenstein}}]{ZHSL}
\bibinfo{author}{\bibfnamefont{K.}~\bibnamefont{{\.Z}yczkowski}},
  \bibinfo{author}{\bibfnamefont{P.}~\bibnamefont{Horodecki}},
  \bibinfo{author}{\bibfnamefont{A.}~\bibnamefont{Sanpera}}, \bibnamefont{and}
  \bibinfo{author}{\bibfnamefont{M.}~\bibnamefont{Lewenstein}},
  \bibinfo{journal}{Phys. Rev. A} \textbf{\bibinfo{volume}{58}},
  \bibinfo{pages}{883} (\bibinfo{year}{1998}).

\bibitem[{\citenamefont{Slater}(2005{\natexlab{a}})}]{slaterJGP}
\bibinfo{author}{\bibfnamefont{P.~B.} \bibnamefont{Slater}},
  \bibinfo{journal}{J. Geom. Phys.} \textbf{\bibinfo{volume}{53}},
  \bibinfo{pages}{74} (\bibinfo{year}{2005}{\natexlab{a}}).

\bibitem[{\citenamefont{Slater}(2005{\natexlab{b}})}]{slaterPRA}
\bibinfo{author}{\bibfnamefont{P.~B.} \bibnamefont{Slater}},
  \bibinfo{journal}{Phys. Rev. A} \textbf{\bibinfo{volume}{71}},
  \bibinfo{pages}{052319} (\bibinfo{year}{2005}{\natexlab{b}}).

\bibitem[{\citenamefont{Slater}({\natexlab{a}})}]{slaterChinese}
\bibinfo{author}{\bibfnamefont{P.~B.} \bibnamefont{Slater}},
  \eprint{quant-ph/0505093}.

\bibitem[{\citenamefont{Szarek et~al.}(2006)\citenamefont{Szarek, Bengtsson,
  and {\.Z}yczkowski}}]{sbz}
\bibinfo{author}{\bibfnamefont{S.}~\bibnamefont{Szarek}},
  \bibinfo{author}{\bibfnamefont{I.}~\bibnamefont{Bengtsson}},
  \bibnamefont{and}
  \bibinfo{author}{\bibfnamefont{K.}~\bibnamefont{{\.Z}yczkowski}},
  \bibinfo{journal}{J. Phys. A} \textbf{\bibinfo{volume}{39}},
  \bibinfo{pages}{L119} (\bibinfo{year}{2006}).

\bibitem[{\citenamefont{Innami}(1999)}]{innami}
\bibinfo{author}{\bibfnamefont{N.}~\bibnamefont{Innami}},
  \bibinfo{journal}{Proc. Amer. Math. Soc.} \textbf{\bibinfo{volume}{127}},
  \bibinfo{pages}{3049} (\bibinfo{year}{1999}).

\bibitem[{\citenamefont{Petz}(1994)}]{petz1994}
\bibinfo{author}{\bibfnamefont{D.}~\bibnamefont{Petz}}, \bibinfo{journal}{J.
  Math. Phys.} \textbf{\bibinfo{volume}{35}}, \bibinfo{pages}{780}
  (\bibinfo{year}{1994}).

\bibitem[{\citenamefont{Gibilisco and Isola}(2003)}]{wigneryanase}
\bibinfo{author}{\bibfnamefont{P.}~\bibnamefont{Gibilisco}} \bibnamefont{and}
  \bibinfo{author}{\bibfnamefont{T.}~\bibnamefont{Isola}}, \bibinfo{journal}{J.
  Math. Phys.} \textbf{\bibinfo{volume}{44}}, \bibinfo{pages}{3752}
  (\bibinfo{year}{2003}).

\bibitem[{\citenamefont{\.Zyczkowski}(2005)}]{mathscinet}
\bibinfo{author}{\bibfnamefont{K.}~\bibnamefont{\.Zyczkowski}},
  \bibinfo{journal}{MathSciNet} \textbf{\bibinfo{volume}{MR2102050}}
  (\bibinfo{year}{2005}).

\bibitem[{\citenamefont{Slater}({\natexlab{b}})}]{pbsJak}
\bibinfo{author}{\bibfnamefont{P.~B.} \bibnamefont{Slater}},
  \eprint{quant-ph/0508227}.

\bibitem[{\citenamefont{Slater}(2006)}]{pbsCanosa}
\bibinfo{author}{\bibfnamefont{P.~B.} \bibnamefont{Slater}},
  \bibinfo{journal}{J. Phys. A} \textbf{\bibinfo{volume}{39}},
  \bibinfo{pages}{913} (\bibinfo{year}{2006}).

\bibitem[{\citenamefont{Slater}(2000)}]{slaterC}
\bibinfo{author}{\bibfnamefont{P.~B.} \bibnamefont{Slater}},
  \bibinfo{journal}{Euro. Phys. J. B} \textbf{\bibinfo{volume}{17}},
  \bibinfo{pages}{471} (\bibinfo{year}{2000}).

\bibitem[{\citenamefont{Livio}(2002)}]{livio}
\bibinfo{author}{\bibfnamefont{M.}~\bibnamefont{Livio}},
  \emph{\bibinfo{title}{The Golden Ratio}} (\bibinfo{publisher}{Broadway},
  \bibinfo{address}{New York}, \bibinfo{year}{2002}).

\bibitem[{\citenamefont{Jak\'obczyk and Siennicki}(2001)}]{jak}
\bibinfo{author}{\bibfnamefont{L.}~\bibnamefont{Jak\'obczyk}} \bibnamefont{and}
  \bibinfo{author}{\bibfnamefont{M.}~\bibnamefont{Siennicki}},
  \bibinfo{journal}{Phys. Lett. A} \textbf{\bibinfo{volume}{286}},
  \bibinfo{pages}{383} (\bibinfo{year}{2001}).

\bibitem[{\citenamefont{Bloore}(1976)}]{bloore}
\bibinfo{author}{\bibfnamefont{F.~J.} \bibnamefont{Bloore}},
  \bibinfo{journal}{J. Phys. A} \textbf{\bibinfo{volume}{9}},
  \bibinfo{pages}{2059} (\bibinfo{year}{1976}).

\bibitem[{\citenamefont{Kimura and Kossakowski}(2005)}]{kk}
\bibinfo{author}{\bibfnamefont{G.}~\bibnamefont{Kimura}} \bibnamefont{and}
  \bibinfo{author}{\bibfnamefont{A.}~\bibnamefont{Kossakowski}},
  \bibinfo{journal}{Open Sys. Inform. Dyn.} \textbf{\bibinfo{volume}{12}},
  \bibinfo{pages}{207} (\bibinfo{year}{2005}).

\bibitem[{\citenamefont{Byrd and Khaneja}(2003)}]{byrd}
\bibinfo{author}{\bibfnamefont{M.~S.} \bibnamefont{Byrd}} \bibnamefont{and}
  \bibinfo{author}{\bibfnamefont{N.}~\bibnamefont{Khaneja}},
  \bibinfo{journal}{Phys. Rev. A} \textbf{\bibinfo{volume}{68}},
  \bibinfo{pages}{062322} (\bibinfo{year}{2003}).

\bibitem[{\citenamefont{Tilma et~al.}(2002)\citenamefont{Tilma, Byrd, and
  Sudarshan}}]{sudarshan}
\bibinfo{author}{\bibfnamefont{T.}~\bibnamefont{Tilma}},
  \bibinfo{author}{\bibfnamefont{M.}~\bibnamefont{Byrd}}, \bibnamefont{and}
  \bibinfo{author}{\bibfnamefont{E.~C.~G.} \bibnamefont{Sudarshan}},
  \bibinfo{journal}{J. Phys. A} \textbf{\bibinfo{volume}{35}},
  \bibinfo{pages}{10445} (\bibinfo{year}{2002}).

\bibitem[{\citenamefont{Mkrtchian and Chaltykyan}(1987)}]{vanik}
\bibinfo{author}{\bibfnamefont{V.~E.} \bibnamefont{Mkrtchian}}
  \bibnamefont{and} \bibinfo{author}{\bibfnamefont{V.~O.}
  \bibnamefont{Chaltykyan}}, \bibinfo{journal}{Opt. Commun.}
  \textbf{\bibinfo{volume}{63}}, \bibinfo{pages}{239} (\bibinfo{year}{1987}).

\bibitem[{\citenamefont{Fano}(1983)}]{fano}
\bibinfo{author}{\bibfnamefont{U.}~\bibnamefont{Fano}}, \bibinfo{journal}{Rev.
  Mod. Phys.} \textbf{\bibinfo{volume}{55}}, \bibinfo{pages}{855}
  (\bibinfo{year}{1983}).

\bibitem[{\citenamefont{Scutaru}(2004)}]{scutaru}
\bibinfo{author}{\bibfnamefont{H.}~\bibnamefont{Scutaru}},
  \bibinfo{journal}{Proc. Romanian Acad., Ser. A} \textbf{\bibinfo{volume}{5}},
  \bibinfo{pages}{1} (\bibinfo{year}{2004}).

\bibitem[{\citenamefont{Kryszewski and Zachcial}()}]{stan}
\bibinfo{author}{\bibfnamefont{S.}~\bibnamefont{Kryszewski}} \bibnamefont{and}
  \bibinfo{author}{\bibfnamefont{M.}~\bibnamefont{Zachcial}},
  \eprint{quant-ph/0602065}.

\bibitem[{\citenamefont{Slater}(1999)}]{slaterA}
\bibinfo{author}{\bibfnamefont{P.~B.} \bibnamefont{Slater}},
  \bibinfo{journal}{J. Phys. A} \textbf{\bibinfo{volume}{32}},
  \bibinfo{pages}{5261} (\bibinfo{year}{1999}).

\bibitem[{\citenamefont{Slater}(2002)}]{slaterqip}
\bibinfo{author}{\bibfnamefont{P.~B.} \bibnamefont{Slater}},
  \bibinfo{journal}{Quant. Info. Proc.} \textbf{\bibinfo{volume}{1}},
  \bibinfo{pages}{397} (\bibinfo{year}{2002}).

\bibitem[{\citenamefont{Slater}({\natexlab{c}})}]{slaterHSPRA}
\bibinfo{author}{\bibfnamefont{P.~B.} \bibnamefont{Slater}},
  \eprint{quant-ph/0609006}.

\bibitem[{\citenamefont{Horn and Johnson}(1985)}]{horn}
\bibinfo{author}{\bibfnamefont{R.~A.} \bibnamefont{Horn}} \bibnamefont{and}
  \bibinfo{author}{\bibfnamefont{C.~A.} \bibnamefont{Johnson}},
  \emph{\bibinfo{title}{Matrix Analysis}} (\bibinfo{publisher}{Cambridge},
  \bibinfo{address}{New York}, \bibinfo{year}{1985}).

\bibitem[{\citenamefont{Brown}(2001)}]{cylindrical}
\bibinfo{author}{\bibfnamefont{C.~W.} \bibnamefont{Brown}},
  \bibinfo{journal}{J. Symbolic Comput.} \textbf{\bibinfo{volume}{31}},
  \bibinfo{pages}{521} (\bibinfo{year}{2001}).

\bibitem[{\citenamefont{Strzebonski}(2002)}]{strzebonski}
\bibinfo{author}{\bibfnamefont{A.}~\bibnamefont{Strzebonski}},
  \bibinfo{journal}{Mathematica Journal} \textbf{\bibinfo{volume}{7}},
  \bibinfo{pages}{10} (\bibinfo{year}{2002}).

\bibitem[{\citenamefont{Peres}(1996)}]{asher}
\bibinfo{author}{\bibfnamefont{A.}~\bibnamefont{Peres}},
  \bibinfo{journal}{Phys. Rev. Lett.} \textbf{\bibinfo{volume}{77}},
  \bibinfo{pages}{1413} (\bibinfo{year}{1996}).

\bibitem[{\citenamefont{Horodecki et~al.}(1996)\citenamefont{Horodecki,
  Horodecki, and Horodecki}}]{michal}
\bibinfo{author}{\bibfnamefont{M.}~\bibnamefont{Horodecki}},
  \bibinfo{author}{\bibfnamefont{P.}~\bibnamefont{Horodecki}},
  \bibnamefont{and}
  \bibinfo{author}{\bibfnamefont{R.}~\bibnamefont{Horodecki}},
  \bibinfo{journal}{Phys. Lett. A} \textbf{\bibinfo{volume}{223}},
  \bibinfo{pages}{1} (\bibinfo{year}{1996}).

\bibitem[{\citenamefont{Bru{\ss} and Macchiavello}(2005)}]{bruss}
\bibinfo{author}{\bibfnamefont{D.}~\bibnamefont{Bru{\ss}}} \bibnamefont{and}
  \bibinfo{author}{\bibfnamefont{C.}~\bibnamefont{Macchiavello}},
  \bibinfo{journal}{Found. Phys.} \textbf{\bibinfo{volume}{35}},
  \bibinfo{pages}{1921} (\bibinfo{year}{2005}).

\bibitem[{\citenamefont{F.~Verstraete and Moor}(2001)}]{ver}
\bibinfo{author}{\bibfnamefont{J.}~\bibnamefont{F.~Verstraete}}
  \bibnamefont{and} \bibinfo{author}{\bibfnamefont{B.~D.} \bibnamefont{Moor}},
  \bibinfo{journal}{Phys. Rev. A} \textbf{\bibinfo{volume}{64}},
  \bibinfo{pages}{010101} (\bibinfo{year}{2001}).

\bibitem[{\citenamefont{Hall}(1998)}]{hall}
\bibinfo{author}{\bibfnamefont{M.~J.~W.} \bibnamefont{Hall}},
  \bibinfo{journal}{Phys. Lett. A} \textbf{\bibinfo{volume}{242}},
  \bibinfo{pages}{123} (\bibinfo{year}{1998}).

\bibitem[{\citenamefont{Macdonald}(1995)}]{ig}
\bibinfo{author}{\bibfnamefont{I.~G.} \bibnamefont{Macdonald}},
  \emph{\bibinfo{title}{Symmetric Functions and Hall Polynomial}}
  (\bibinfo{publisher}{Oxford}, \bibinfo{address}{New York},
  \bibinfo{year}{1995}).

\bibitem[{\citenamefont{Hansen}()}]{hansen}
\bibinfo{author}{\bibfnamefont{F.}~\bibnamefont{Hansen}},
  \eprint{math-ph/0601056}.

\bibitem[{\citenamefont{Audenaert et~al.}()\citenamefont{Audenaert,
  Calsamiglia, Masanes, Munoz-Tapia, Acin, Bagan, and Verstraete}}]{chernoff}
\bibinfo{author}{\bibfnamefont{K.~M.~R.} \bibnamefont{Audenaert}},
  \bibinfo{author}{\bibfnamefont{J.}~\bibnamefont{Calsamiglia}},
  \bibinfo{author}{\bibfnamefont{L.}~\bibnamefont{Masanes}},
  \bibinfo{author}{\bibfnamefont{R.}~\bibnamefont{Munoz-Tapia}},
  \bibinfo{author}{\bibfnamefont{A.}~\bibnamefont{Acin}},
  \bibinfo{author}{\bibfnamefont{E.}~\bibnamefont{Bagan}}, \bibnamefont{and}
  \bibinfo{author}{\bibfnamefont{F.}~\bibnamefont{Verstraete}},
  \eprint{quant-ph/0610027}.

\bibitem[{\citenamefont{Nielsen}(2000)}]{nielsen}
\bibinfo{author}{\bibfnamefont{M.}~\bibnamefont{Nielsen}},
  \bibinfo{journal}{Phys. Rev. A} \textbf{\bibinfo{volume}{61}},
  \bibinfo{pages}{064301} (\bibinfo{year}{2000}).

\bibitem[{\citenamefont{Gurvits and Barnum}(2002)}]{sepsize1}
\bibinfo{author}{\bibfnamefont{L.}~\bibnamefont{Gurvits}} \bibnamefont{and}
  \bibinfo{author}{\bibfnamefont{H.}~\bibnamefont{Barnum}},
  \bibinfo{journal}{Phys.Rev. A} \textbf{\bibinfo{volume}{66}},
  \bibinfo{pages}{062311} (\bibinfo{year}{2002}).

\bibitem[{\citenamefont{Lockhart}(2002)}]{lockhart}
\bibinfo{author}{\bibfnamefont{R.}~\bibnamefont{Lockhart}},
  \bibinfo{journal}{Phys. Rev. A} \textbf{\bibinfo{volume}{65}},
  \bibinfo{pages}{064304} (\bibinfo{year}{2002}).

\bibitem[{\citenamefont{Szarek}(2005)}]{sz1}
\bibinfo{author}{\bibfnamefont{S.}~\bibnamefont{Szarek}},
  \bibinfo{journal}{Phys. Rev. A} \textbf{\bibinfo{volume}{72}},
  \bibinfo{pages}{032304} (\bibinfo{year}{2005}).

\bibitem[{\citenamefont{Aubrun and Szarek}()}]{sz2}
\bibinfo{author}{\bibfnamefont{G.}~\bibnamefont{Aubrun}} \bibnamefont{and}
  \bibinfo{author}{\bibfnamefont{S.}~\bibnamefont{Szarek}},
  \eprint{quant-ph/0503221}.

\bibitem[{\citenamefont{Gurvits and Barnum}(2003)}]{gb1}
\bibinfo{author}{\bibfnamefont{L.}~\bibnamefont{Gurvits}} \bibnamefont{and}
  \bibinfo{author}{\bibfnamefont{H.}~\bibnamefont{Barnum}},
  \bibinfo{journal}{Phys. Rev. A} \textbf{\bibinfo{volume}{68}},
  \bibinfo{pages}{042310} (\bibinfo{year}{2003}).

\bibitem[{\citenamefont{Gurvits and Barnum}(2005)}]{gb2}
\bibinfo{author}{\bibfnamefont{L.}~\bibnamefont{Gurvits}} \bibnamefont{and}
  \bibinfo{author}{\bibfnamefont{H.}~\bibnamefont{Barnum}},
  \bibinfo{journal}{Phys. Rev. A} \textbf{\bibinfo{volume}{72}},
  \bibinfo{pages}{032322} (\bibinfo{year}{2005}).

\bibitem[{\citenamefont{Hildebrand}()}]{hildebrand}
\bibinfo{author}{\bibfnamefont{R.}~\bibnamefont{Hildebrand}},
  \eprint{quant-ph/0601201}.

\bibitem[{\citenamefont{Ulrich and Watson}(1994)}]{ulrich}
\bibinfo{author}{\bibfnamefont{G.}~\bibnamefont{Ulrich}} \bibnamefont{and}
  \bibinfo{author}{\bibfnamefont{L.~T.} \bibnamefont{Watson}},
  \bibinfo{journal}{SIAM J. Sci. Comput.} \textbf{\bibinfo{volume}{15}},
  \bibinfo{pages}{528} (\bibinfo{year}{1994}).

\bibitem[{\citenamefont{Wang}()}]{wang}
\bibinfo{author}{\bibfnamefont{A.~M.} \bibnamefont{Wang}},
  \eprint{quant-ph/0002073}.

\bibitem[{\citenamefont{Fortuna et~al.}(2003)\citenamefont{Fortuna, Gianni,
  Parenti, and Traverso}}]{fortuna}
\bibinfo{author}{\bibfnamefont{E.}~\bibnamefont{Fortuna}},
  \bibinfo{author}{\bibfnamefont{P.}~\bibnamefont{Gianni}},
  \bibinfo{author}{\bibfnamefont{P.}~\bibnamefont{Parenti}}, \bibnamefont{and}
  \bibinfo{author}{\bibfnamefont{C.}~\bibnamefont{Traverso}},
  \bibinfo{journal}{J. Symb. Comput.} \textbf{\bibinfo{volume}{36}},
  \bibinfo{pages}{343} (\bibinfo{year}{2003}).

\end{thebibliography}

\end{document}